\documentclass[aps,prb,twocolumn,showpacs,superscriptaddress]{revtex4}

\usepackage{graphicx}
\usepackage{amsmath}
\usepackage{amssymb}
\usepackage{lscape}
\usepackage{braket}
\usepackage{float,hyperref,epstopdf}
\DeclareMathOperator{\Tr}{Tr}

\begin{document}
\title{Spontaneous magnetization of quantum $XY$ spin model in joint presence of\\
 quenched and annealed disorder}

\author{Anindita Bera}
\affiliation{Department of Applied Mathematics, University of Calcutta, 92, A.P.C. Road, Kolkata 700 009, India}
\affiliation{Harish-Chandra Research Institute, Chhatnag Road, Jhunsi, Allahabad 211 019, India}
\author{Debraj Rakshit}
\affiliation{Harish-Chandra Research Institute, Chhatnag Road, Jhunsi, Allahabad 211 019, India}

\author{Aditi Sen(De)}
\affiliation{Harish-Chandra Research Institute, Chhatnag Road, Jhunsi, Allahabad 211 019, India}
\author{Ujjwal Sen}
\affiliation{Harish-Chandra Research Institute, Chhatnag Road, Jhunsi, Allahabad 211 019, India}

\date{\today}
\begin{abstract}
We investigate equilibrium statistical properties of the quantum $XY$ spin-1/2 model
in an external magnetic field 
%at equilibrium
 when %the system parameters 
the interaction 	and field parts 
are subjected to quenched or/and annealed disorder. The randomness present in the system are termed annealed or quenched depending on the relation between two different time scales - 
%quenched disorder corresponds to the situation when 
the time scale associated with the equilibriation of the randomness 
%is large compared to
and the time of observation.
 Within a mean-field framework, we study the effects of disorders on spontaneous magnetization,
 both by perturbative and numerical techniques.
Our primary interest is to understand the differences between quenched and annealed cases, and also to investigate the interplay when both of them are present in a system. 
We observe in particular
%that in all cases when there is a symmetry-breaking disordered field, irrespective of whether it is 
%quenched or annealed, and irrespective of other particulars of the governing Hamiltonian,
%when the quenched and annealed randomness are introduced in the interaction and field terms, 
%the system magnetizes in the direction parallel or transverse to the applied field. Interestingly, 
that 
%in the case 
when interaction and field terms are respectively quenched and annealed, 
%both perturbative analysis and numerical simulation confirm that 
critical temperature for the system to magnetize in the direction parallel to the applied field does not depend on any of the disorders. 
Further, an annealed 
disordered interaction neither affects the magnetizations nor the critical temperatures. We carry out a comparative study 
of the different combinations of the disorders in the interaction and field terms, and point out their generic features.
% that
% exclusively 
%depend upon the specific nature of the disorders.
\end{abstract}

\maketitle

\section{Introduction}
\label{sec_introduction}

Disorder is an unavoidable feature of condensed matter and atomic many-body systems, in both classical as well as quantum domains  \cite{disoredr-review} . There are long-standing quests to understand several non-trivial quantum phenomena caused due to the presence of disorder. Some of the prominent examples in the quantum case are  disorder induced localizations \cite{anderson,anderson2, loc2}, high $T_c$ superconductivity \cite{Auerbach}, and novel quantum phases \cite{mezard,sachdev,yao}.

For disordered parameters in a physical system, two typical situations may arise depending on the interrelation between the two fundamental time scales associated with the disordered parameter, viz. the time scale over which the disorder configuration equilibriates and the time scale associated with observation of the physical quantities of interest \cite{disoredr-review,mezard, anqu-1,anqu-2,anqu-3,brout}. For the cases where the system's disorder configuration remains effectively frozen throughout the entire observation process, the disorder is considered to be ``quenched". In such cases, one obtains the quenched averaged free energy as the logarithm of the partition function, averaged over several realizations of the random parameters. However, there can be a separate situation where the observation takes place during the equilibriation process of the disorder parameters, so that the time scale associated with the configurational change is of the order or within a few orders of the observation time. In such cases, the randomness of the corresponding system parameters should be considered to be ``annealed" and the partition function has to be averaged over several random realizations, so that the annealed averaged free energy is obtained by taking the logarithm of the averaged partition function.

%Quantum spin models have drawn a lot of attention in recent times as they mimic a wide range of physical systems. Randomness occurs naturally in the system parameters of the physical realizations of quantum spin models. 

There has been continuous efforts to understand effects of disorder in quantum systems \cite{ahufinger,Nagaoka,wehr-1,wehr-2,universal,group}. A particular reason of recent interest in such systems is also due to the fact that current technology allows us to realize artificial randomness in a controlled way, in for example ultracold atoms trapped in optical lattices \cite{ahufinger}. A considerable amount of effort has been dedicated to investigate disordered systems that are quenched \cite{wehr-1,wehr-2,group, universal,loc2}, and in particular to understand the effects of such disorder on the universal dynamics in the vicinity of quantum phase transitions \cite{universal}, a useful test for which is given by the Harris Criterion \citep{harris}. Unlike quenched disorder, the universality class of a phase transition is usually not affected by the presence of annealed disorder, as the partition function after averaging over random realizations can be replaced by one corresponding to an effective model which is free from the disorder. In other instances, quenched disorder spin systems have been studied to understand ``glassy" properties in type II superconductors \cite{Rosenstein}, to demonstrate break-down of thermalization in the presence of disorder \cite{loc2}, to achieve quantum advantages due to the introduction of the disorder \cite{group} and to explore disorder induced quantum phenomena such as ``order from disorder" \cite{group,odd-dis}. Significant works have also been carried out for studying the consequences of annealed disorder in the spin systems as well \cite{Thorpe,hide}. Moreover, efforts have been directed towards understanding system properties at equilibrium when the nature of the disorder changes from quenched to annealed \cite{anqu-3}.

In this respect, an important question, which to our knowledge is yet to be dealt with, is how the thermodynamical quantities respond in joint presence of quenched and annealed disorders. It is for example possible to inquire about the properties of a quantum
spin system governed by the Hamiltonian \(\mathcal{H} = \mathcal{H}_{int} + \mathcal{H}_{field}\), 
where the disorder introduced in one set of parameters, say the couplings in  \(\mathcal{H}_{int}\), 
remain quenched during the observation process, while the equilibriating time scales of another set, say the field parameters in
\(\mathcal{H}_{field}\), is of same or near order as the observation time scale, so that the latter colection
 forms an annealed set of parameters. The main focus of this paper is to study 
the equilibrium properties, such as magnetization and critical temperatures, of
such systems, and compare between them and with systems having only quenched or only annealed disorder or systems devoid of disorder.

Spontaneous magnetization in higher-dimensional quantum $XY$ model in presence of an unidirectional quenched random field has been considered with a lot of interest in recent times. It has been shown that spontaneous magnetization perishes when a small random magnetic field with appropriate symmetry is introduced in the $XY$ spin systems \cite{wehr-1,imry,imbrie}. However, 
%spontaneous magnetization 
it 
persists in absence of the appropriate symmetry of the external random field \cite{wehr-2,classical}. Interestingly, it has been shown that a uniaxial random field may help the system to magnetize even in two-dimension \cite{wehr-2,Crawford}. Recently, mean-field approach\cite{mf general, mf interaction} has been adapted to look into the aspects of spontaneous magnetizations and critical scalings in quenched disordered spin models \cite{classical}. 

In this work, we investigate spontaneous magnetization subjected to random interaction or/and random external field in disordered quantum spin-$1/2$ $XY$ models, with a mean-field approximation. The interaction is annealed disordered or quenched disordered or ordered. The transverse magnetic field is again chosen from these three options. We 
compare between systems having 
%consider 
all the nine possible combinations of interaction and field with respect to their disorder.
%, and compare between the different situations. 
Presence of randomness in the interaction strength preserves isotropic symmetry of the system, while 
%. However, an 
a small random field, even with zero mean, breaks the same
 and the system magnetizes in either parallel or perpendicular direction to the applied random field. We derive analytical expressions for the critical temperatures and near-critical magnetizations for all the cases. Our analysis reveals, for example, that annealed disorder present in the interaction term does not have any effect on the system's spontaneous magnetization and the corresponding critical temperature, irrespective of the presence or absence of quenched or annealed disorder in the field. However, annealed disorder considered in the field term affects the transverse magnetization while keeping the magnetization parallel to the applied field unaltered. Quenched randomness always causes the magnetizations to shrink in value, whether or not there is an accompanying annealed disorder in another parameter of the system. However, we find that there can be situations where the critical temperature is not affected by the presence of quenched disorder.

The rest of the paper is arranged as follows. In Sec.~\ref{anqu}, we present a general recapitulation of the mechanism for obtaining the annealed and the quenched averaged values of physical observables. In Sec.~\ref{treatment}, we introduce the system and its mean field treatment. We also discuss about the various situations depending on the nature of the disorder parameters and 
the segment of the Hamiltonian in which the disorder is located. 
%nature and station
In Sec.~\ref{example}, we present a detailed analysis for a quantum spin-1/2 model in joint presence of quenched and annealed disorders. Section \ref{other} tabulates the analytical expressions for the critical temperatures and scalings of the magnetizations near the critical points for the different types of disorder in the quantum spin-1/2 model. We conclude in Sec.~\ref{summary}.\\

\section{Annealed and quenched disorder}
\label{anqu}
In this section, we briefly discuss the mechanism for computing the annealed and quenched averaged values of the observables. As mentioned earlier, the distinction between quenched and annealed disorder is determined by relative comparison of two different time scales of the physical system under consideration, viz. the relaxation time associated with the equilibriation of the disorders, say $\tau_1$, and the time necessary for the required observation on the system,  say $\tau_2$. For the cases where $\tau_1$ is of same or near order of magnitude of $\tau_2$, the statistical properties of the system at equilibrium is obtained via annealed averaging, 
which is calculated by averaging of the partition function, $\mathcal{Z}$, over several random realizations. The free energy for systems with the annealed disorder is given as $\mathcal{F}=-(1/\beta)\ln\langle \mathcal{Z} \rangle$, where $\beta=1/(\kappa_B \mathcal{T})$ with $\kappa_B$ being the Boltzmann constant and $\mathcal{T}$ being the absolute temperature. Here, and in the rest of the paper the notation $\langle . \rangle$ shall imply an average of the argument over the relevant disorder degrees of freedom. However, if $\tau_1 \gg \tau_2$, i.e., the impurities remain trapped in random but fixed positions during the observation time, the statistical properties of the system at equilibrium is obtained via quenched averaging. In case of quenched averaging, the logarithm of the partition function, instead of the partition function itself, is averaged over several random realizations. The free energy in presence of the quenched disorder is given by $\mathcal{F}=-(1/\beta)\langle \ln \mathcal{Z} \rangle$. Let us note here that for both types of disorders as well as for the ordered systems, the observation time is assumed to be much longer than the relaxation of the spin degrees of freedom.

Let us consider a general Hamiltonian $\mathcal{H} = \mathcal{H}\left(\{a_i\},\{q_j\}\right)$, where $\{a_i\}$ and $\{q_j\}$ are two sets of system parameters that are respectively annealed and quenched disordered. We introduce the functional partition function $\mathcal{Z}\left(\{a_i\},\{q_j\},\{\lambda_k\}\right)=\Tr\left[ e^{ -\beta \left\{   \mathcal{H}\left( \{a_i\},\{q_j\} \right) +\sum_k \lambda_k \mathcal{A}_k \right\} } \right]$, where the term $\sum_k \lambda_k \mathcal{A}_k$ is an auxiliary function. The auxiliary function is used later for obtaining the expectation values of the operator $\mathcal{A}_i$ by taking derivatives with respect to $\lambda_i$ at $\lambda_k=0$ $\forall k$, where  $\mathcal{Z}$ has been implicitly assumed to be a differentiable function of the $\lambda_k$'s. The functional free energy, $\mathcal{F}$,  after performing configurational averaging over the two types of disorder, reads as 
\begin{widetext}
\begin{eqnarray}
\mathcal{F}(\{\lambda_k\}) = -\frac{1}{\beta}\int \prod_j d{q_j} \mathcal{P}_j (q_j) \ln \left\{ \int \prod_i {d}a_i \mathcal{P}_i (a_i) \mathcal{Z}\left(\{a_i\},\{q_j\},\{\lambda_k\}\right) \right\}, 
\end{eqnarray}
\end{widetext}
where $\mathcal{P}_i (a_i)$ and $\mathcal{P}_i (q_i)$ represents the probability density functions of the annealed and quenched parameters, respectively. The thermodynamic average of the observable $\mathcal{A}_k$, averaged also over the disorder degrees of freedom, can finally be obtained as $\partial \mathcal{F}/\partial \lambda_k|_{\lambda_i=0} \forall i$.

\section{System Hamiltonian and mean field treatment}
\label{treatment}
We investigate the isotropic quantum $XY$ spin model in an external magnetic field with disorder in the interaction part or in the field part or in both, within a mean-field approximation. We are primarily interested in drawing a comparative analysis on the effect of disorder on spontaneous magnetization and their scalings near the critical point as a function of temperature, with different possible combinations of disorder. For example, a possible combination is quenched disorder in the coupling terms and annealed disorder in the field terms. In this section, we introduce the system and its mean-field treatment.

The general form of the Hamiltonian of the ferromagnetic quantum $XY$ model in presence of disorder in both the interaction and the coupling parts is given by $\mathcal{H}_{XY}({\tilde{\eta}_{ij}},{\eta_i}) = \mathcal{H}_{int}({\tilde{\eta}_{ij}})+\mathcal{H}_{ext}({\eta_i})$, where 
\begin{eqnarray}
\mathcal{H}_{int} ({\tilde{\eta}_{ij}}) &=&-\sum_{(i, j)\in S} (\mathcal{J'}+\tilde{\epsilon} \tilde{\eta}_{ij}) [\sigma_x^i \sigma_{x}^j+\sigma_y^i \sigma_{y}^j],\nonumber \\
\mathcal{H}_{ext}({\eta_i})&=&-\epsilon \sum_{i=1}^\mathcal{N} \eta_i \sigma_y^i.
\end{eqnarray}
Here the coupling constant $\mathcal{J'}>0$. The indices, $i$ and $j$, denote the sites of an arbitrary $d$-dimensional lattice and  $\sigma_i^\alpha, \alpha=x, y$, are the Pauli matrices at the $i^{th}$ site. $\mathcal{N}$ is the total number of spins. The set $S$ denotes a subset of the set of all (unordered) pairs of lattice sites. Both $\tilde{\epsilon}$ and $\epsilon$ are non-negative parameters, namely the dimension of energy, that quantify the strengths of the corresponding random parameters. The unidirectional random field is chosen to be directed along the $y$-axis. $\tilde{\eta}_{ij}$ and $\eta_{i}$ are independent and identically distributed (dimensionless) Gaussian random variables with zero mean and variance $f$. The constant $f$ is a dimensionless quantity that depends on the Hamiltonian and the lattice on which it is defined. We shall discuss further about it in the next paragraph.
 
In the mean-field approach, we approximate the interaction term by $-\sum_{(i,j)\in S} (\mathcal{J'}+\tilde{\epsilon} \tilde{\eta}_{ij}) (m_x \sigma_x^i+m_y \sigma_{y}^i)$, where $m_x$ and $m_y$ are the spins as well as disorder averaged magentizations of the system at absolute temperature $\mathcal{T}$. $m_x$ and $m_y$ are therefore mean-field variables to be obtained from the self-consistency equations of the mean-field theory. The interaction term can be further rewritten as $-\sum_{i=1}^\mathcal{N} (\mathcal{J}+\tilde{\epsilon} \tilde{\eta}_{i}) (m_x \sigma_x^i+m_y \sigma_{y}^i)$, where $\mathcal{J}=\mathcal{J'} f$ and $\tilde{\eta}_{i}$ are Gaussian random variables with zero mean and unit variance. Note therefore that $f$ is the number of different $j$'s for a given $i$ in the set $S$. For nearest-neighbour interactions in one-dimension, $f=1$, while for the same in the two dimensional square lattice, $f=2$.
% every spin is assumed to interact with all other spins (not just with the nearest-neighbour). Within this approximation, for large number of spins $\mathcal{N}$, the interaction part approximately reads $\mathcal{H}_{int} (\tilde{\eta})=-(1/N)[\sum_{\alpha}\sum_{j;j\ne i}^N (\mathcal{J}+\tilde{\epsilon} \tilde{\eta}_i) \sigma_i^{\alpha}]\sigma_{j}^{\alpha}=-\sum_{\alpha} (\mathcal{J}+\tilde{\epsilon} \tilde{\eta}) m_{\alpha} \sigma_i^{\alpha}$,  where $m_{\alpha}= (1/\mathcal{N}) \sum_{j=1}^{\mathcal{N}} \sigma_{j}^{\alpha}$. 
Hence, within the mean-field  approximation, the Hamiltonian, $\mathcal{H}_{XY}$, can be written as
\begin{equation}
\label{xy_disorder1}
\mathcal{H}_{XY}(\tilde{\eta},\eta)=-(\mathcal{J}+\tilde{\epsilon} \tilde{\eta}) (m_x \sigma_x+m_y \sigma_y)-\epsilon \eta \sigma_y.
\end{equation}
Note that in Eq.~(\ref{xy_disorder1}), the Hamiltonian corresponding to the ordered system can be obtained by simply setting $\tilde{\epsilon}=\epsilon=0$. The quantities $\tilde{\epsilon}$ and $\epsilon$ are chosen to be small compared to $\mathcal{J}$.
The functional partition function of the system in the canonical equilibrium state at absolute temperature $\mathcal{T}$ is given by
\begin{equation}
\mathcal{Z}\left(\tilde{\eta},\eta, \{\lambda_k\}\right)=\Tr\left[e^{-\beta\left\{ \mathcal{H}_{XY} \left(\tilde{\eta},\eta\right) + \sum_{k} \lambda_k \mathcal{A}_k \right\}}\right].
\end{equation}
%where $\sum_{\alpha} \mathcal{B}_{\alpha} \sigma_{\alpha}$, as suggested in the previous section, is the auxiliary functional to be used for obtaining the components of the magnetization, $m_{\alpha}$, through functional derivatives with respect to $\mathcal{B}_{\alpha}$. 
Note that for the cases, where disorder is present in either the interaction part or the field part, the functional partition function reduces to $\mathcal{Z}\left(\tilde{\eta}, \{\lambda_k\}\right)$ $\left[\mathcal{Z}\left(\eta, \{\lambda_k\}\right)\right]$ for $\epsilon=0$ $\left[\tilde{\epsilon}=0\right]$.

Now let us consider three different categories: 
\begin{itemize}
\item[] \emph{Category (i):} Both the interaction as well as the field terms are annealed disordered, or any of them is so, while the other is ordered. 
\item[] \emph{Category (ii):} Both the interaction and the field terms are quenched disordered, or any one of them is so, while the other is ordered. 
\item[] \emph{Category (iii):} The interaction and field terms are respectively quenched and annealed disordered or vice-versa.
\end{itemize}
% $(i)$ Both the interaction and the field terms are annealed disorder, either one of the terms is annealed disorder and the second one is free from any disorder, $(ii)$ both the interaction and the field parts are quenched disorder, either one of the terms is quenched disorder and the second one is ordered, and $(iii)$ One of the terms is quenched disordered, whereas the second one is annealed disordered. 
As mentioned earlier, for the cases within the first category, the free energy is obtained by performing a disorder average over the partition function, whereas for the cases within the second category, the free energy is obtained by performing a disorder average over the logarithm of the partition function. For the cases within the third category, first a configurational averaging of the partition function over the annealed parameters for fixed realization of quenched randomness is performed. This is followed by quenched averaging of the logarithm of the annealed averaged partition function in order to obtain the final result. As an example, let us consider a quantum spin magnetic system in presence of quenched randomness in the parameter associated with the interaction part, $\langle \tilde{\eta} \rangle$, and annealed randomness in the parameter associated with the field part $\langle \eta \rangle$. The functional free energy for this case is obtained as 
\begin{widetext}
\begin{equation}
\label{an-qu-XY}
\mathcal{F}(\{\lambda_k\}) = -\frac{1}{\beta(2 \pi \sqrt{\Delta \tilde{\Delta}})}\int_{-\infty}^{\infty} d\eta~e^{-\frac{\eta^2}{2 \Delta}} \ln \left\{ \int_{-\infty}^{\infty} d \tilde{\eta}~e^{-\frac{\tilde{\eta}^2}{2 \tilde{\Delta}}} \Tr\left[e^{-\beta\left\{ \mathcal{H}_{XY} \left(\tilde{\eta},\eta\right) + \sum_{k} \lambda_k \mathcal{A}_k \right\}}\right]\right\},
\end{equation}
\end{widetext}
where $\tilde{\eta}$ and $\eta$ are assumed to be independent Gaussian random variables with vanishing mean and standard deviations $\tilde{\Delta}$ and $\Delta$ respectively.
%where $\Delta_1$ and $\Delta_2$ are the widths of distribution each one of which assumes the value of unity throughout this work. 
For convenience, we alternatively represent Eq.~(\ref{an-qu-XY}) as 
\begin{equation}
\mathcal{F}(\{\lambda_k\}) = -\frac{1}{\beta} \left\langle \ln \left\langle \Tr\left[e^{-\beta\left\{ \mathcal{H}_{XY} \left(\tilde{\eta},\eta\right) + \sum_{k} \lambda_k \mathcal{A}_k \right\}}\right] \right\rangle_{\eta}\right\rangle_{\tilde{\eta}}.
\end{equation}
We shall assume that the disordered parameters are drawn from independent Gaussian distributions with zero mean and unit variance.
% The $x$- and the $y$-component of magnetization are obtained as 
%\begin{equation}
%\label{m_alpha}
%m_{\alpha}=\frac{\partial\mathcal{F}}{\partial \mathcal{B}_{\alpha}}|_{\{B_x=0,B_y=0\}}.
%\end{equation}
In the following section, we present a detailed analysis of the spontaneous magnetization and critical scalings of the quantum spin-1/2 spin model for this representative case.

\section{Quantum XY spin-1/2 model in Joint presence of Quenched and Annealed disorder}
\label{example}
We now investigate 
%In this section we shall discuss 
the behavior of spontaneous magnetizations of the isotropic quantum spin-$1/2$ $XY$ model in the presence of both quenched and annealed disorders. This corresponds to the category (iii) of the preceding section. Our system Hamiltonian is given by  Eq.~(\ref{xy_disorder1}). The field and interaction parts are subjected to annealed and quenched disorders, respectively. Starting from  Eq.~(\ref{an-qu-XY}) and following straightforward algebraic steps, the components of magnetization along the $x$- and $y$-axes can be obtained by solving for common zeros of the following pair of functions:
\begin{widetext}
\begin{eqnarray}
\label{fx}
f_x^{\tilde{\epsilon},\epsilon}(\vec{m}) = \left\langle \left[{\left\langle\cosh(\beta k)\right\rangle_\eta}\right]^{-1} \left\langle {m_x (\mathcal{J}+\tilde{\epsilon} \tilde{\eta}) \sinh(\beta k)}/{k}\right\rangle_{\eta} \right\rangle_{\tilde{\eta}}-m_x,
\end{eqnarray}
and 
\begin{eqnarray}
\label{fy}
f_y^{\tilde{\epsilon},\epsilon}(\vec{m})  = \left\langle \left[{\left\langle\cosh(\beta k)\right\rangle_\eta}\right]^{-1} \left\langle {\left[m_y \left(\mathcal{J}+\tilde{\epsilon} \tilde{\eta}\right)+\epsilon \eta \right] \sinh(\beta k)}/{k}\right\rangle_{\eta}\right\rangle_{\tilde{\eta}}
-m_y,
\end{eqnarray}
\end{widetext}
where $k=\sqrt{\left[m_x \left(\mathcal{J}+\tilde{\epsilon} \tilde{\eta}\right)\right]^2+\left[m_y \left(\mathcal{J}+\tilde{\epsilon} \tilde{\eta}\right)+\epsilon \eta\right]^2}$ and $\vec{m}=(m_x,m_y)$.
We perform a perturbative analysis for solving the coupled set of equations formed by equating the functions $f_x^{\tilde{\epsilon},\epsilon}$ and $f_y^{\tilde{\epsilon},\epsilon}$ to zero. The perturbative approach helps us to derive the exact analytical expressions for the near-critical temperature and scaling of magnetization. Moreover, we carry out numerical analysis, which helps us to look into effects of disorder in the system properties as functions of temperature, in near-critical as well as far-from-critical regimes. 

\subsection{Critical point and scaling of magnetization near criticality}
\label{near}
When strengths of the random parameters are small, it turns out that perturbative analyses yield a great deal of insight about the system's behavior. Such analyses, in particular, provide quantitative values of critical temperatures and near-critical scalings of magnetization. Bivariate Taylor series expansions of Eqs.~(\ref{fx}) and (\ref{fy}) around $\epsilon/\mathcal{J}$  and $\tilde{\epsilon}/\mathcal{J}$ at $\epsilon=0$ and $\tilde{\epsilon}=0$ give the leading order  behaviors of $f_{x}^{\tilde{\epsilon},\epsilon}(\vec{m})$ and $f_{y}^{\tilde{\epsilon},\epsilon}(\vec{m})$ as
\begin{equation}
\label{fx12}
f_{x}^{\tilde{\epsilon},\epsilon}(\vec{m})=a_{x}+\frac{1}{2} \mathcal{J}^2 b_{x} \left(\frac{\tilde{\epsilon}}{\mathcal{J}}\right)^2+\frac{1}{2} \mathcal{J}^2 c_{x} \left(\frac{\epsilon}{\mathcal{J}}\right)^2 + \cdots,
\end{equation}
and 
\begin{equation}
\label{fy12}
f_{y}^{\tilde{\epsilon},\epsilon}(\vec{m})=a_{y}+\frac{1}{2} \mathcal{J}^2 b_{y} \left(\frac{\tilde{\epsilon}}{\mathcal{J}}\right)^2+\frac{1}{2} \mathcal{J}^2 c_{y} \left(\frac{\epsilon}{\mathcal{J}}\right)^2 + \cdots,
\end{equation}
where
\begin{equation}
a_x=\frac{m_x}{m} \tanh[\beta \mathcal{J} m]-m_x,
\end{equation}
\begin{equation}
a_y = \frac{m_y}{m} \tanh[\beta \mathcal{J} m]-m_y,
\end{equation}
\begin{eqnarray}
b_x =\frac{-2 \beta^2 m_x m \tanh[\beta \mathcal{J} m]}{\cosh[\beta \mathcal{J} m]^2} ,
\end{eqnarray}
\begin{eqnarray}
b_y = \frac{-2 \beta^2 m_y m \tanh[\beta \mathcal{J} m]}{\cosh[\beta \mathcal{J} m]^2},
\end{eqnarray}
\begin{widetext}
\begin{eqnarray}
c_x = \frac{m_x m_y^2}{\mathcal{J} m^4} \left[\frac{3 \tanh[\beta \mathcal{J} m]}{\mathcal{J} m}+\beta (\tanh[\beta \mathcal{J} m]^2-3)\right]+\frac{m_x}{\mathcal{J} m^2} \left[\frac{\beta}{\cosh[\beta \mathcal{J} m]^2}-\frac{\tanh[\beta \mathcal{J} m]}{\mathcal{J} m}\right],
\end{eqnarray}
\begin{eqnarray}
c_y =\frac{m_x^2 m_y}{\mathcal{J} m^4} \left[-\frac{3 \tanh[\beta \mathcal{J} m]}{\mathcal{J} m}+\beta (3-\tanh[\beta \mathcal{J} m]^2)\right],
\end{eqnarray}
\end{widetext}
with $m=|\vec{m}|=\sqrt{m_x^2+m_y^2}$.

\begin{figure}[h!]
%\vspace*{+.2cm}
\includegraphics[angle=0, width=65mm]{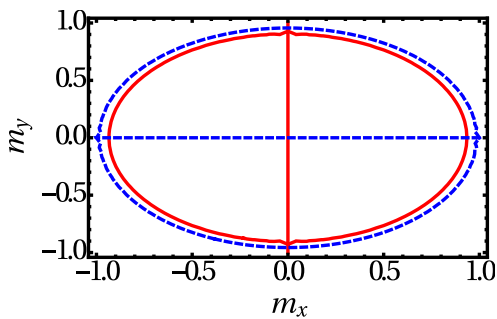}
%%%\includegraphics[angle=0, width=85mm]{xy_contourplot.eps}
%\vspace*{0.2cm}
\caption{(Color online.) 
Contour plot showing directions of magnetization in presence of the disorder. Zero contour lines corresponding to $f_x^{\tilde{\epsilon},\epsilon}(\vec{m})$ and $f_y^{\tilde{\epsilon},\epsilon}(\vec{m})$ in Eqs.~(\ref{fx12}) [solid-red] and (\ref{fy12}) [dotted-blue] for $\epsilon/\mathcal{J}=0.2$, $\tilde{\epsilon}/\mathcal{J}=0.05$, and  $\mathcal{J} \beta=2$, as functions of $m_x$ and $m_y$. All quantities are dimensionless.}
\label{fig_contour}
\end{figure}

The ordered system with vanishing $\tilde{\epsilon}$ and $\epsilon$ has a continuous (circular) symmetry, which implies that magnetization behaves uniformly in all possible directions. The continuous symmetry of the system is broken in presence of the unidirectional annealed disorder. The possible directions of magnetizations can  be deduced forthwith via a contour analysis \cite{classical}. This is done by identifying the zero-contour lines corresponding to the functions $f_{x}^{\tilde{\epsilon},\epsilon}(\vec{m})$ and $f_{y}^{\tilde{\epsilon},\epsilon}(\vec{m})$ (see Eqs.~(\ref{fx12}) and (\ref{fy12})), and
 the intersection points of the 
 %zero-contour 
 lines are solutions of the magnetization.
See Fig.~\ref{fig_contour}.
% \#\#\#. 
  Contour analysis suggest two possible solutions: The system magnetizes either in the transverse direction of the external annealed field , i.e., $m_x \neq 0, m_y = 0$ (case I) or in the parallel direction of the random field, i.e., $m_x = 0, m_y \neq 0$ (case II). By setting $\vec{m}=(m \cos [\phi], m \sin [\phi])$, the transverse and the parallel magnetization correspond to $\phi=0$ and $\pi/2$ respectively. For ease of reference, we will henceforth 
%We shall interchangably 
use $m_\perp$ for $m_x$ ($m_\parallel$ for $m_y$) to refer to the transverse (parallel) magnetization.

In order to derive the expressions for the critical temperature and the scalings of the magnetizations near criticality, we perform another round of Taylor expansions in Eqs.~(\ref{fx12}) and (\ref{fy12}) around $m=0$. The leading order behavior of the functions $f_{x}^{\tilde{\epsilon},\epsilon}(\vec{m})$ and $f_{y}^{\tilde{\epsilon},\epsilon}(\vec{m})$ for small $m$ are given by
\begin{eqnarray}
\label{fx3}
f_x^{\tilde{\epsilon},\epsilon} (\vec{m})=\left[-1+\mathcal{J} \left(\beta-\frac{\epsilon^2 \beta^3}{3}\right)\right] m \cos[\phi]+
\nonumber\\
\frac{1}{3!} \left[\frac{2}{5} \mathcal{J} \beta^3 (\mathcal{J}^2 (-5+4 \epsilon^2 \beta^2)-15 \tilde{\epsilon}^2)\right] m^3 \cos[\phi]^3+O(m^5)
\nonumber\\
\end{eqnarray}
and
\begin{eqnarray}
\label{fy3}
f_y^{\tilde{\epsilon},\epsilon} (\vec{m})=(-1+\mathcal{J} \beta) m \sin[\phi]+
\nonumber\\
\frac{1}{3!} \left[-2 J \beta^3 \left(J^2+3 \tilde{\epsilon}^2\right)\right] m^3 \sin[\phi]^3+O(m^5).
\end{eqnarray}
Now as discussed earlier in context of the contour analysis, the allowed values of $\phi$ are $0$ (case I) and $\pi/2$ (case II). For transverse magnetization, $\phi=0$ and $f_y^{\tilde{\epsilon},\epsilon} (\vec{m})$ vanishes identically. Eq.~(\ref{fx3}) leads us to
\begin{equation}
\label{m1}
m_\perp = \pm \sqrt{5} \sqrt{\frac{3 (\mathcal{J} \beta -1)- \mathcal{J} \epsilon^2 \beta^3}{\mathcal{J} \beta^3 \left[\mathcal{J}^2 \left(5-4 \epsilon^2 \beta^2\right)+15 \tilde{\epsilon}^2\right]}}.
\end{equation}
%We shall interchangably use $m_\perp$ and $m_x$ ($m_\parallel$ and $m_y$) for the transverse (parallel) magnetization.
The system magnetizes in the perpendicular direction only below a certain critical temperature. This can be obtained by setting $m_\perp =0$, whence the critical temperature is given by
\begin{equation}
\label{b1}
\beta_{c,\perp}= \frac{1}{\mathcal{J}}+\frac{\epsilon^2}{3 \mathcal{J}^3}.
\end{equation}

Similarly, the parallel magnetization can be obtained by setting $\phi=\pi/2$ 
in  Eqs.~(\ref{fx3}) and (\ref{fy3}). For this case, the expression in Eq.~(\ref{fx3}) vanishes identically, and by equating the expression in Eq.~(\ref{fy3}) to zero, we obtain 
\begin{equation}
\label{m2}
m_\parallel = \pm \sqrt{3} \sqrt{\frac{\mathcal{J} \beta-1}{\mathcal{J} \beta^3 (\mathcal{J}^2+3 \tilde{\epsilon}^2)}}.
\end{equation}
Setting $m_\parallel=0$, we find the critical temperature to be given by 
\begin{equation}
\label{b2}
\beta_{c,\parallel}= \frac{1}{\mathcal{J}}.
\end{equation}
Note that from the set of Eqs.~(\ref{m1}) -- (\ref{b2}), one can recover the results for the ordered system by setting $\tilde{\epsilon} = 0$ and $\epsilon=0$. We find that for  case being studied in this section, i.e., for a system with quenched randomness in the interaction term and annealed randomness in the field term, both parallel and transverse magnetizations survive the onslaught of the defects in the system as modelled by the disordered parameters in the Hamiltonian. 
Interestingly, the critical temperatures are not affected by the presence of the quenched disorder (in the interaction terms). Moreover, the annealed randomness in the field terms does not influence the parallel critical temperature, although it lowers the transverse critical temperature. Our analysis also reveals that both transverse and parallel magnetizations are lowered in magnitude compared to the ordered system due to the presence of quenched disorder. However, it is only the transverse magnetization on which the annealed randomness has an impact, and the effect is to reduce the magnetization, while
the parallel magnetization remains unfazed in presence of the annealed disorder in the field term.

\begin{figure}[t]
%\vspace*{+.9cm}
\includegraphics[angle=0,width=6cm,height=4.2cm]{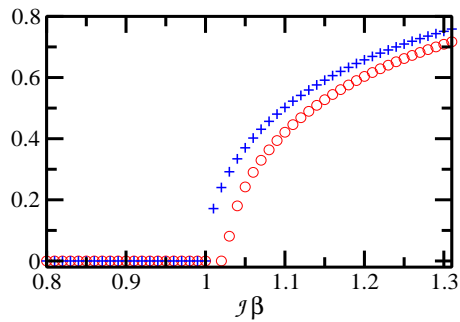}
%\includegraphics[angle=0,width=7cm,height=4cm]{p1.eps}
%\vspace*{0.2cm}
\caption{(Color online.) 
%Persistence of spontaneous magnetization in the joint presence of quenched and annealed disorder. 
Spontaneous transverse magnetization in the joint presence of quenched and annealed disorder.
The plot shows numerical results for transverse magnetization 
%in specific directions 
in presence of annealed disorder in the field term and quenched disorder in the interaction term, and is compared to the same in the pure system.
% where no disorder present in both the field and interaction terms i.e. $\epsilon/\mathcal{J}=0$ and $\tilde{\epsilon}/\mathcal{J}=0$. 
%Magnetization is plotted as a function of $J \beta$, in the direction transverse to the disordered field . 
Blue pluses correspond to the magnetization of the pure system and red circles correspond to the transverse magnetization in presence of annealed disorder in the field term and quenched disorder in the interaction term, obtained by solving for roots of  Eqs.~(\ref{fx}) and (\ref{fy}) with $\epsilon/\mathcal{J}=0.1$ and $\tilde{\epsilon}/\mathcal{J}=0.15$.
%The blue dots correspond to the roots of  with $\epsilon/\mathcal{J}=0.1$ and $\tilde{\epsilon}/\mathcal{J}=0.1$. 
All quantities are dimensionless. The vertical axis represents the transverse magnetization for the disordered case, and 
the magnetization in the pure case.}
\label{fig_epsilon}
\end{figure}

\subsection{Away from critical point}
Away from the critical point, the perturbative approach fails. We numerically find out the roots of the coupled set of equations,
obtained by setting the expressions in Eqs.~(\ref{fx}) and (\ref{fy}) equal to zero, i.e., \(f_x^{\tilde{\epsilon},\epsilon}(\vec{m}) =0\) and \(f_y^{\tilde{\epsilon},\epsilon}(\vec{m}) =0\). We perform the configurational averaging  for 8000 random realizations for each type of disorder, viz., $\eta$ and $\tilde{\eta}$. As 
%it may now be expected from our analytics, 
predicted by the perturbative approach,
the numerical simulations also indicate two possible directions of magnetization -- the system can either magnetize along the transverse direction of the external annealed disordered field or it can magnetize in the direction parallel to it. 

In Fig.~\ref{fig_epsilon}, we show the results obtained from numerical analysis for the transverse magnetization, i.e., $m_\perp$ is non-zero and $m_\parallel=0$
% (case I), 
for $\epsilon/\mathcal{J}=0.1$ and $\tilde{\epsilon}/\mathcal{J}=0.15$. At high temperature (above the critical temperature), the system does not magnetize. For $\beta > \beta_{c,\perp}$,  the system magnetizes in the direction which is tranverse to the applied random field. For the ordered system, the critical temperature corresponds to $J \beta=1$. The spontaneous magnetization persists in the presence of disorder, albeit with a reduced critical temperature. The behavior of parallel magnetization, $m_{||}$, obtained from 
the numerical simulations also confirms the trends from the perturbative derivations.

%However, the critical temperature decreases. From our peturbative analysis, we understand by now that the shift in the critical temperature is only due to the quenched randomness in the interaction term and annealed randomness in the field as no effect on the critical temperature. 
% An example of case II, i.e., vanishing $x$-component and non-zero $y$-component, is shown in Fig.~\ref{fig_epsilon}(b) for same choices of $\epsilon$ and $\tilde{\epsilon}$. We find that the critical point, $\beta_{c,\parallel}$, does not depend on the disorder parameters. We also find that the numerical analysis is consistent with the perturbative derivations.\\

\section{Other combinations of quenched and annealed disorders}
\label{other}

To perform comparative studies between different kinds of disordered systems, we shall now adopt similar techniques 
as in the preceding section. 
%
%In this section,  we 
Let us now consider all possible combinations of the three categories as mentioned in Sec. \ref{treatment}, obtained by considering different types of disorders 
 in the interaction and the field terms of the Hamiltonian in Eq.~(\ref{xy_disorder1}). One of these cases has already been discussed 
 in the last section.  
%Our aim is to build a complete understanding on the system's behavior depending on the specific kinds of disorders under consideration. 
We summarize our results in Table 1 that considers all such possible combinations and also includes cases where disorder is absent. We introduce the following notations for convenience: ${\langle \eta \rangle}_{a}$ (${\langle \tilde{\eta} \rangle}_{a}$) implies that the disorder in the field (interaction) term is annealed. Moreover, the same symbol also denotes the mean of the corressponding annealed distribution. On the other hand, ${\langle \eta \rangle}_{q}$ (${\langle \tilde{\eta} \rangle}_{q}$) implies that the disorder in the field (interaction) term is quenched, and the same symbol also denotes the mean of the corresponding quenched distribution. The variances of the distributions of all the disordered random variables are taken to be unity.
We also use following shorthand notation: $a_1=\mathcal{J} \beta-1$, $a_2=3 (1-\mathcal{J} \beta)+\mathcal{J} \epsilon^2 \beta^3$, $a_3=1-\mathcal{J} \beta + \mathcal{J} \beta^3 \epsilon^2$, $b_1=-5+4 \epsilon^2 \beta^2$, $b_2=4 \epsilon^2 \beta^2-1$, $b_3 = \mathcal{J}^2+3 \tilde{\epsilon}^2$. 

\begin{widetext}
\begin{center}
\begin{table}[h]
\begin{tabular}{|c|c|c|c|c|c|c|}
\hline
  Case  &  interaction term & field term & $m_\perp$ & $m_\parallel$ & $\beta_{c,\perp}$ & $\beta_{c,\parallel}$ \\ \hline \hline

1  &  $\tilde{\eta}=0$ & ${\eta}=0$  & $\pm \sqrt{3} \sqrt{\frac{a_1}{\mathcal{J}^3 \beta^3}}$     &  $\pm \sqrt{3} \sqrt{\frac{a_1}{\mathcal{J}^3 \beta^3}}$  & $\frac{1}{\mathcal{J}}$  & $\frac{1}{\mathcal{J}}$    \\ \hline

2  &  ${\langle \tilde{\eta} \rangle}_{a}=0$ & ${\eta}=0$ & $\pm \sqrt{3} \sqrt{\frac{a_1}{\mathcal{J}^3 \beta^3}}$   &  $\pm \sqrt{3} \sqrt{\frac{a_1}{\mathcal{J}^3 \beta^3}}$  & $\frac{1}{\mathcal{J}}$  & $\frac{1}{\mathcal{J}}$      \\ \hline

3  &  $\tilde{\eta}=0$ &  ${\langle {\eta} \rangle}_{a}=0$   & $\pm \sqrt{5} \sqrt{\frac{a_2}{b_1 \mathcal{J}^3 \beta^3}}$     & $\pm \sqrt{3} \sqrt{\frac{a_1}{\mathcal{J}^3 \beta^3}}$  & $\frac{1}{\mathcal{J}}+\frac{\epsilon^2}{3 \mathcal{J}^3}$  & $\frac{1}{\mathcal{J}}$    \\ \hline

4  & ${\langle \tilde{\eta} \rangle}_{a}=0$  & ${\langle {\eta} \rangle}_{a}=0$    & $\pm \sqrt{5} \sqrt{\frac{a_2}{b_1 \mathcal{J}^3 \beta^3}}$   &  $\pm \sqrt{3} \sqrt{\frac{a_1}{\mathcal{J}^3 \beta^3}}$  &   $\frac{1}{\mathcal{J}}+\frac{\epsilon^2}{3 \mathcal{J}^3}$  & $\frac{1}{\mathcal{J}}$      \\ \hline

5  & ${\langle \tilde{\eta} \rangle}_{q}=0$  & ${\eta}=0$  & $\pm \sqrt{3} \sqrt{\frac{a_1}{b_3 \mathcal{J} \beta^3}}$    & $\pm \sqrt{3} \sqrt{\frac{a_1}{b_3 \mathcal{J} \beta^3}}$  & $\frac{1}{\mathcal{J}}$   &  $\frac{1}{\mathcal{J}}$       \\ \hline

6  &   $\tilde{\eta}=0$ & ${\langle {\eta} \rangle}_{q}=0$  & $\pm \sqrt{5} \sqrt{\frac{a_2}{b_1 \mathcal{J}^3 \beta^3}}$    & $\pm \sqrt{3} \sqrt{\frac{a_3}{b_2 \mathcal{J}^3 \beta^3}}$  & $\frac{1}{\mathcal{J}}+\frac{\epsilon^2}{3 \mathcal{J}^3}$  &  $\frac{1}{\mathcal{J}}+\frac{\epsilon^2}{\mathcal{J}^3}$   \\ \hline

7  & ${\langle \tilde{\eta} \rangle}_{q}=0$ & ${\langle {\eta} \rangle}_{q}=0$   &  $\pm \sqrt{5} \sqrt{\frac{a_2}{\mathcal{J} \beta^3 [b_1 \mathcal{J}^2-15 \tilde{\epsilon}^2]}}$   &  $\pm \sqrt{3} \sqrt{\frac{a_3}{\mathcal{J} \beta^3 (b_2 \mathcal{J}^2-3 \tilde{\epsilon}^2)}}$   &  $\frac{1}{\mathcal{J}}+\frac{\epsilon^2}{3 \mathcal{J}^3}$  &   $\frac{1}{\mathcal{J}}+\frac{\epsilon^2}{\mathcal{J}^3}$  \\ \hline

8  &  ${\langle \tilde{\eta} \rangle}_{a}=0$ & ${\langle {\eta} \rangle}_{q}=0$  & $\pm \sqrt{5} \sqrt{\frac{a_2}{b_1 \mathcal{J}^3 \beta^3}}$    & $\pm \sqrt{3} \sqrt{\frac{a_3}{b_2 \mathcal{J}^3 \beta^3}}$  & $\frac{1}{\mathcal{J}}+\frac{\epsilon^2}{3 \mathcal{J}^3}$  &  $\frac{1}{\mathcal{J}}+\frac{\epsilon^2}{\mathcal{J}^3}$   \\ \hline

9  &  ${\langle \tilde{\eta} \rangle}_{q}=0$ & ${\langle {\eta} \rangle}_{a}=0$ & $\pm \sqrt{5} \sqrt{\frac{a_2}{\mathcal{J} \beta^3 [b_1 \mathcal{J}^2-15 \tilde{\epsilon}^2]}}$    & $\pm \sqrt{3} \sqrt{\frac{a_1}{b_3 \mathcal{J} \beta^3}}$  & $\frac{1}{\mathcal{J}}+\frac{\epsilon^2}{3 \mathcal{J}^3}$   & $\frac{1}{\mathcal{J}}$        \\ \hline

\end{tabular}
\caption{ A comparison of the magnetizations and the critical temperatures for the different combinations of disorders. $m_\perp$ and $m_\parallel$ denotes, respectively, the magnetizations tranverse and parallel to the applied random field. $\beta_{c,\perp}$ and $\beta_{c,\parallel}$ are proportional to the inverse of the critical temperatures in the transverse and parallel directions respectively.
}
\label{table:combinations}
\end{table}
\end{center}
\end{widetext}

Here we briefly describe the results summarized in Table I. Case 1 corresponds to the case when the system is free from any kind of disorder. 
%This is a purely isotropic case, where t
The isotropic quantum $XY$ model in absence of any external field manifests a spontaneous magnetization which has a continuous circular symmetry. The spontaneous magnetization occurs below a critical temperature constrained by the condition $\beta =1/\mathcal{J}$. 

Let us now discuss the cases that belong to category ($i$), as described in Sec.~\ref{treatment}. 
As expected, the system retains its circular symmetry in presence of an annealed disorder in the interaction part (case 2). Surprisingly, the presence of an annealed disorder in the interaction part neither disturbs the magnitude of the  magnetization nor does it shift the critical temperature. However, when the clean system is subjected to an annealed randomness, only in the field term (case 3), the situation changes. The circular symmetry of the system is broken and the system now possesses magnetization in the direction either parallel or transverse to the applied field. Although the magnitude of the parallel magnetization, $m_{\parallel}$, and corresponding critical temperature, $(\kappa_B \beta_{c,\parallel})^{-1}$, remain unaltered due to this annealed field, the magnitude of $m_{\perp}$ as well as the corresponding critical temperature are lowered compared to the ordered system. 
We find that the results in case 4, where annealed disorder is present both in the  interaction and the field terms, 
are identical with those
% the results for which are identical to those 
of case 3. 
This is intuitively understandable from our  analyses in cases 1 and 2, where the presence of an annealed disorder in the interaction term has no efect on the magnetizations and critical temperatures of the system in the mean field limit. 

Let us now look into the cases that belongs to category ($ii$) of Sec. \ref{treatment}. Case 5 represents the situation when there is quenched randomness in the interaction term. The system preserves continuous symmetry of the spontaneous magnetizations,
%. Presence of quenched randomness in the interaction does not perturb 
the critical temperature remains unaltered. However, the magnetization gets affected and shrinks in magnitude. For the case where interaction is ordered but the field is quenched disordered (case 6), the continuous symmetry is broken, the system exhibits transverse and parallel magnetizations, albeit with a lowered value compared to the magnetization in the clean system and
%. Also, the system magnetizes at 
requires
lower temperatures to magnetize.
% in comparison to the clean system. 
Interestingly the effect of disorder is more pronounced in the parallel direction than in the transverse direction. Finally, we find that the behaviour of the system with quenched randomness in both interaction and field parts (case 7) is qualitatively similar to the previous case.

Finally, we consider the cases in category (\(iii\)) of Sec. \ref{treatment}. For the cases in this category, 
annealed and quenched disorders are both introduced in the system -- one in the interaction term and another in the field term. 
One of such scenarios (case 9)  was considered at length in Sec.\ref{example}.
The other one is case 8. Consistent with what  we have seen in previous cases, the annealed disordered interaction 
does not have any effect 
on the magnetizations and the critical temperatures. Any disorder effect in this case is only due to the quenched disorder present in the field term. Therefore, the perturbative formulae in this case are identical with case 6, where there was no disorder
present in the interaction term.

%we consider annealed and quenched disorders in the interaction and field terms, respectively. The annealed disorder in the interaction term has no effect in both transverse and parallel directions, so that $m_\perp$ and $m_\parallel$ are the same as in case VI, where no disorder is present in the interaction term. In case IX, we consider quenched and annealed disorders in the interaction and field terms, respectively. 
% We find that annealed disorder has effect only in the transverse magnetization and does not have any impact the parallel magnetization. For this reason, $m_\perp$ in this case IX is same with the case VII. 

%So From Table (\ref{table:combinations}), we observe that annealed disorder in the interaction term given in (\ref{xy_disorder1}) has no effect in the transverse magnetization as well as parallel magnetization.  Also annealed disorderd in the field term has no effect in the parallel magnetization.

\section{conclusions}
\label{summary}
In summary, this work examines quantum spin-1/2 $XY$ models with continuous and broken continuous isotropic symmetries within the mean field framework, and investigates the effect on spontaneous magnetization due to the presence of disorders in external field or/and in the couplings. The disorders we consider can be annealed or quenched in nature. 

%We present a generic formalism for investing the statistical properties of such systems at equilibrium depending on the type of disorders. 
A combined approach of perturbative analysis and numerical simulation have been adopted for characterizing the spontaneous magnetization in the systems.  We derive exact analytical expressions, within a perturbative approach,  for the critical temperatures and near-critical scalings of magnetization corresponding to the various combinations of the disorders, and carry out a comparative study. The results obtained within the perturbative theory is found to match with those obtained from the numerical simulations. We find that spontaneous magnetization persists in the presence of randomness in these  models. The ordered system and the disordered systems with randomness only in couplings exhibit magnetization for all possible orientations due to the continuous circular symmetry. The circular symmetry breaks down in presence of an infinitesimal unidirectional disordered field. In presence of the random field, which can be annealed or quenched, the system exhibits magnetization for two selective orientations -- parallel or transverse to the external field.  Interestingly, annealed disorder present in the interaction term does not have any effect on the critical temperatures as well as the magnetizations of the system. However, an annealed random field perturbs the transverse magnetization although the parallel magnetization remains unaltered. Unlike the annealed one, presence of quenched disorder always shrinks the value of magnetization in the system, 
%and the reduction, for small disorder strength, 
%the reduction in magnetization magnitude 
which 
is proportional to the square of the disorder strength in the perturbative analyses. A key focus has been on systems that exhibit a joint presence of annealed and quenched disorders, and we discuss the corresponding effect on spontaneous magnetization and its critical temperature.

\acknowledgments
A.B. acknowledges the support of the Department of Science and Technology (DST), Government of India, through the award of an INSPIRE fellowship.


\begin{thebibliography}{100}

\bibitem{disoredr-review} K. Binder and A. P. Young, Rev. Mod. Phys. {\bf58}, 801 (1986); D. Belitz, T. R. Kirkpatrick, and T. Vojta, \emph{ibid.} {\bf77}, 579 (2005); A. Das and B. K. Chakrabarti, \emph{ibid.} {\bf80}, 1061 (2008); H. Alloul, J. Bobroff, M. Gabay, and P. J. Hirschfeld, \emph{ibid.} {\bf81}, 45 (2009).


\bibitem{anderson}
P. W. Anderson, \emph{Basic Notions of Condensed Matter Physics} (Westview Press, Colorado, 1984); P. A. Lee and T. V. Ramakrishnan, Rev. Mod. Phys. {\bf57}, 287 (1985); 
R. Zallen, \emph{The physics of amorphous solids} (Wiley, New York, 1998).

 \bibitem{anderson2}
P. W. Anderson, Phys. Rev. {\bf109}, 1492 (1958); 
E. Abrahams, P. W. Anderson, D. C. Licciardello, and T. V. Ramakrishnan, Phys. Rev. Lett. {\bf42}, 673 (1979).


\bibitem{loc2} A. Pal and D. A. Huse, Phys. Rev. B {\bf82}, 174411 (2010); M. Znidaric, T. Prosen, and P. Prelovsek, Phys. Rev. B {\bf77}, 064426 (2008); E. Canovi, D. Rossini, R. Fazio, G. E. Santoro, and A. Silva, Phys. Rev. B {\bf83}, 094431 (2011); J. H. Bardarson, F. Pollmann, and J. E. Moore, Phys. Rev.
Lett. {\bf109}, 017202 (2012); J. Eisert, M. Friesdorf, and C. Gogolin, Nat. Phys. {\bf11},
124 (2015).


\bibitem{Auerbach} A. Auerbach, \emph{Interacting electrons and Quantum magnetism} (Springer, New York, 1994).

\bibitem{mezard} D. Chowdhury, \emph{Spin Glasses and other Frustrated Systems} (Wiley, New York, 1986);  M. Mezard, G. Parisi, and M. A. Virasoro, \emph{Spin Glass Theory and Beyond} (World Scientific, Singapore, 1987). 

\bibitem{sachdev} S. Sachdev, \emph{Quantum Phase Transitions} (Cambridge University Press, Cambridge, 1999).

\bibitem{yao} Z. Yao, K. P. C. da Costa, M. Kiselev, and N. Prokof'ev, Phys. Rev. Lett. {\bf112}, 225301 (2014); J. P. {\'A}. {\'Z}{\~u}niga and N. Laflorencie, \emph{ibid.} {\bf111}, 160403 (2013).

\bibitem{anqu-1} L.F. Cugliandolo, {\it Disordered systems}, Lecture notes, (Cargese, 2011); S. G. Abaimov, {\it Statistical Physics of Non-Thermal Phase Transitions}, (Springer, 2015).

\bibitem{anqu-2}S. M.-Araghi and M. Sebtosheikh, Phys. Rev. E {\bf 92}, 022116 (2015).

\bibitem{anqu-3} F. P. de A. Prado and G. M. Sch{\"u}tz, J. Stat. Phys. {\bf 142}, 984 (2011); A. N. M.-Kakkada, O. T. Valls, and C. Dasgupta, 	Phys. Rev. B {\bf 90}, 024202 (2014). 

\bibitem{brout} R. Brout, Phys. Rev. {\bf 115}, 824 (1959).

\bibitem{ahufinger} V. Ahufinger, L. S.-Palencia, A. Kantian, A. Sanpera, and M. Lewenstein, Phys. Rev. A {\bf72}, 063616 (2005); M. Lewenstein, A. Sanpera, V. Ahufinger, B. Damski, A. Sen(De), and U. Sen, Adv. Phys. {\bf56}, 243 (2007); L. Fallani, C. Fort, and M. Inguscio, Adv. At. Mol. Opt. Phys. {\bf56}, 119 (2008); A. Aspect and M. Inguscio, Physics Today {\bf62}, 30 (2009); L. S.-Palencia and M. Lewenstein, Nat. Phys. {\bf6}, 87 (2010); G. Modugno, Rep. Prog. Phys. {\bf73}, 102401 (2010); B. Shapiro, J. Phys. A {\bf45}, 143001 (2012);  M. Lewenstein, A. Sanpera, and V. Ahufinger, \emph{Ultracold atoms in Optical Lattices: simulating quantum many body physics} (Oxford University Press, Oxford, 2012).

\bibitem{Nagaoka}
Y. Nagaoka and H. Fukuyama (Eds.), \emph{Anderson Localization}, Springer Series in Solid State Sciences {\bf39}, (Springer, Heidelberg, 1982); T. Ando and H. Fukuyama (Eds.), 
\emph{Anderson Localization}, Springer Proceedings of Physics {\bf28}, (Springer, Heidelberg, 1988).

\bibitem{universal} J. Dziarmaga, Phys. Rev. B {\bf 74}, 064416 (2006); T. Caneva, R. Fazio and G. E. Santoro, Phys. Rev. B {\bf 76}, 144427 (2007).

\bibitem{group}
R. Prabhu, S. Pradhan, A. Sen (De), and U. Sen, Phys. Rev. A {\bf84}, 042334 (2011);
U. Mishra, D. Rakshit, R. Prabhu, A. Sen(De) and U. Sen, arXiv:1408.0179; D. Sadhukhan, S. Singha Roy, D. Rakshit, A. Sen(De), and U. Sen New J. Phys. {\bf 17}, 043013 (2015); D. Sadhukhan, S. Singha Roy, D. Rakshit, R. Prabhu, A. Sen(De), and U. Sen Phys. Rev. E {\bf 93}, 012131 (2016).

\bibitem{wehr-1}
M. Aizenman and J. Wehr, Phys. Rev. Lett. {\bf62}, 2503 (1989); M. Aizenman and J. Wehr, Comm. Math. Phys. {\bf130}, 489 (1990).

\bibitem{wehr-2} J. Wehr, A. Niederberger, L. S.-Palencia, and M. Lewenstein, Phys. Rev. B {\bf74}, 224448 (2006).

\bibitem{harris} A. B. Harris, J. Phys. C {\bf 7}, 1671 (1974).

\bibitem{Rosenstein} B. Rosenstein and D. Li, Rev. Mod. Phys. \textbf{82}, 109 (2010).


\bibitem{odd-dis} A. Aharony, Phys. Rev. B {\bf18}, 3328 (1978); D. E. Feldman, J. Phys. A {\bf31}, L177 (1998); B. J. Minchau and R. A. Pelcovits, Phys. Rev. B {\bf32}, 3081 (1985); I. A. Fomin, J. Low Temp. Phys. {\bf134}, 97 (2005); I. A. Fomin, JETP Lett. {\bf 85}, 434 (2007); see also G. E. Volovik, JETP Lett. {\bf81}, 647 (2005); D. A. Abanin, P. A. Lee, and L. S. Levitov Phys. Rev. Lett. {\bf98}, 156801 (2007); G. E. Volovik, J. Low Temp. Phys. {\bf150}, 453 (2008); A. Niederberger, T. Schulte, J. Wehr, M. Lewenstein, L. Sanchez-Palencia, and K. Sacha, Phys. Rev. Lett. {\bf100}, 030403 (2008); S. Lellouch, T.-L. Dao, T. Koffel, and L. S.-Palencia, Phys. Rev. A {\bf88}, 063646 (2013); A. Niederberger, J. Wehr, M. Lewenstein and K. Sacha, EPL {\bf86}, 26004 (2009); R. L. Greenblatt, M. Aizenman, and J. L. Lebowitz, Phys. Rev. Lett. {\bf103}, 197201 (2009); A. Niederberger, M. M. Rams, J. Dziarmaga, F. M. Cucchietti, J. Wehr, and M. Lewenstein, Phys. Rev. A {\bf82}, 013630 (2010); A. Niederberger, B. Malomed, and M. Lewenstein, Phys. Rev. A {\bf82}, 043622 (2010); P. Lugan and L. S.-Palencia, Phys. Rev. A {\bf84}, 013612 (2011); M. Aizenman, R. L. Greenblatt, and J. L. Lebowitz, J. Math. Phys. {\bf53}, 023301 (2012).

\bibitem{Thorpe} M. F. Thorpe and D. Beeman, Phys. Rev. B {\bf 14}, 188 (1976); H. Falk, J. Phys. C {\bf 9}, L213 (1976); M.F. Thorpe, J. Phys. C {\bf 11}, 2983 (1978).

\bibitem{hide} J. Hide, W. Son, and V. Vedral, Phys. Rev. Lett. {\bf 102}, 100503 (2009); J. Hide, J. Phys. A {\bf 45}, 115302 (2012).


\bibitem{imry} Y. Imry and S. Ma, Phys. Rev. Lett. {\bf35}, 1399 (1975).

\bibitem{imbrie} J. Z. Imbrie, Phys. Rev. Lett. {\bf53}, 1747 (1984); J. Bricmontand A. Kupiainen, \emph{ibid.} {\bf59}, 1929 (1987).

\bibitem{classical} A. Bera, D. Rakshit, M. Lewenstein, A. Sen (De), U. Sen, J. Wehr, Phys. Rev. B {\bf90}, 174408 (2014); A. Bera, D. Rakshit, M. Lewenstein, A. Sen (De), U. Sen, J. Wehr, arXiv: 1509.00704.

\bibitem{Crawford} N. Crawford, J. Stat. Phys. {\bf142}, 11 (2011). See also A. C. D. van Enter, C. K{\" u}lske, O. Christof, A. Alex, and W. M. Ruszel, Braz. J. Prob. Stat. {\bf24}, 226 (2010); N. Crawford, EPL {\bf102}, 36003 (2013); N. Crawford, Comm. Math. Phys. {\bf328}, 203 (2014).

\bibitem{mf general} J. M. Ziman, \emph{Principles of the Theory of Solids} (Cambridge University Press, Cambridge, 1972); K. Huang, \emph{Statistical Mechanics} (Wiley, New York, 1963); N. W. Ashcroft and N. D. Mermin, \emph{Solid State Physics} (Holt, Rinehert and Winston, New York, 1976); R. K. Pathria, \emph{Statistical Mechanics} (Pergamon, New York, 1972); G. D. Mahan, \emph{Many-Particle Physics} (Kluwer/Plenum, New York, 2000);  X.-G. Wen, \emph{Quantum Field Theory of Many-Body Systems} (OUP, Oxford, 2004); L. S.-Palencia, D. Cl{\'e}ment, P. Lugan, P. Bouyer, G. V. Shlyapnikov, and A. Aspect, Phys. Rev. Lett. {\bf 98}, 210401 (2007); L. S.-Palencia, D. Cl{\'e}ment, P. Lugan, P. Bouyer, and A. Aspect, New J. Phys. {\bf 10}, 045019 (2008).

\bibitem{mf interaction} G. Parisi, Lett. in Math. Phys. {\bf 91}, 255 (2009); I. Kondor, J. Phys. A {\bf 16}, L127 (1983); G. Parisi, J. Phys. A: Math. Gen. {\bf 13} 1887 (1980); G. Parisi, arXiv:0706.0094; M. Berciu and R. N. Bhatt, Physica B {\bf 312} 815 (2002).
%
%\bibitem{ek} L. Radzihovsky and A. T. Dorsey, Phys. Rev. Lett. \textbf{88}, 216802 (2002).
%
%
%\bibitem{paanch} Y. Barlas, K. Yang, and A. H. MacDonald, Nanotechnology {\bf 23}, 052001 (2012).
%
%
%\bibitem{tin} D. A. Abanin and D. A. Pesin,Phys. Rev. Lett. {\bf 106}, 136802 (2011).
%
%
%\bibitem{char} L.A. Wray, S. Xu, Y. Xia, D. Qian, A.V. Fedorov, H. Lin, A. Bansil, Y.S. Hor, R.J. Cava, M.Z. Hasan, Nat. Phys. {\bf 6}, 855 (2010).
%
%
%\bibitem{dui} A. Lipi nska, C. Simserides, K. N. Trohidou, M. Goryca, P. Kossacki, A. Majhofer,
%and T. Dietl, Phys. Rev. B {\bf 79},  235322 (2009).
%
%\bibitem{chhoi} M. S. Foster and E. A. Yuzbashyan, Phys. Rev. Lett. {\bf 109},
%246801 (2012).
%
%
%\bibitem{sath} T. R. Kirkpatrick and D. Belitz, Phys. Rev. Lett. {\bf 113},
%127203 (2014).
%
%\bibitem{jordon}
%E. Lieb, T. Schultz, and D. Mattis, Ann. Phys. {\bf 16}, 407 (1961); E. Barouch, B. McCoy, and M. Dresden, Phys. Rev. A {\bf 2}, 1075 (1970); E. Barouch and B. McCoy, ibid. {\bf 3}, 786 (1971). 


%
%\bibitem{Mermin}
%D. Mermin and H. Wagner, Phys. Rev. Lett. {\bf17}, 1133 (1966); 
%P.C. Hohenberg, Phys. Rev. {\bf158}, 383 (1967).

%
%
\end{thebibliography}
\end{document}